\documentclass[11.5pt]{article}
\usepackage[lmargin=25mm,rmargin=35mm,tmargin=35mm, bmargin=35mm, lines=25]{geometry}
\usepackage[utf8]{inputenc}
\usepackage{bm}
\usepackage{amsmath}
\usepackage{pgfplots}
\usepackage{setspace}
\usepackage{tikz}
\usepackage{atbegshi} 
\usepackage{subcaption}
\usepackage{natbib}
\usepackage{lineno}
\usepackage{lipsum}
\setcitestyle{authoryear,open={(},close={)}} 

\DeclareMathOperator*{\argmax}{argmax}

\pgfplotsset{compat=1.17}

\usetikzlibrary {shapes ,arrows , positioning }
\usetikzlibrary{decorations.fractals}
\usetikzlibrary{decorations.pathreplacing}
\usetikzlibrary{arrows.meta}
\usetikzlibrary{automata,chains}

\title{Bayesian Counterfactual Prediction Models for HIV Care Retention with Incomplete Outcome and Covariate Information}

\author{Arman Oganisian$^{1}$, Joseph Hogan$^{1}$, Edwin Sang$^{5}$, \\
Allison DeLong$^{3}$, Ben Mosong$^{5}$, Hamish Fraser$^{2}$, Ann Mwangi$^{4,5}$ \\ 
\vspace{.1in} \\
$^{1}$Department of Biostatistics, Brown University, USA\\
$^{2}$Department of Health Services, Policy, and Practice, Brown University, USA\\
$^{3}$Center for Statistical Sciences, Brown University, USA\\
$^{4}$Department of Mathematics, Physics and Computing, Moi University, Kenya\\
$^{5}$Academic Model Providing Access to Healthcare (AMPATH), Kenya}

\date{}

\begin{document}

\maketitle

\begin{abstract} \singlespacing
Like many chronic diseases, human immunodeficiency virus (HIV) is managed over time at regular clinic visits. At each visit, patient features are assessed, treatments are prescribed, and a subsequent visit is scheduled. There is a need for data-driven methods for both predicting retention and recommending scheduling decisions that optimize retention. Prediction models can be useful for estimating retention rates across a range of scheduling options. However, training such models with electronic health records (EHR) involves several complexities. First, formal causal inference methods are needed to adjust for observed confounding when estimating retention rates under counterfactual scheduling decisions. Second, competing events such as death preclude retention, while censoring events render retention missing. Third, inconsistent monitoring of features such as viral load and CD4 count lead to covariate missingness. This paper presents an all-in-one approach for both predicting HIV retention and optimizing scheduling while accounting for these complexities. We formulate and identify causal retention estimands in terms of potential return-time under a hypothetical scheduling decision. Flexible Bayesian approaches are used to model the observed return-time distribution while accounting for competing and censoring events and form posterior point and uncertainty estimates for these estimands. We address the urgent need for data-driven decision support in HIV care by applying our method to EHR from the Academic Model Providing Access to Healthcare (AMPATH) - a consortium of clinics that treat HIV in Western Kenya.
\end{abstract}

\newpage

\section{Introduction}

The Human Immunodeficiency Virus (HIV), like many diseases, is managed over time with successive followup clinic visits. The HIV Care Continuum \citep{WHO} is a set of public health guidelines for helping patients living with HIV to move from diagnosis to viral suppression. A key component of this continuum is retention in care characterized by ``a person living with HIV who is enrolled in HIV care routinely attending services in accordance with the need'' \citep{UNAIDS2}. Retention is associated with better health outcomes whereas long gaps in care may prevent timely treatments and potentially breed resistance to antiviral therapy. There is therefore strong interest in predicting when patients will miss their scheduled visits both within public policy and clinical communities. At the policy level, the Joint United Nations Programme on HIV/AIDS (UNAIDS) has developed the 95-95-95 \citep{UNAIDS} plan, which sets out targets for having 95\% of patients with HIV diagnosed, 95\% retained in care, and 95\% achieve viral suppression. Individual countries modify these guidelines to set concrete retention definitions. For instance, Kenyan Ministry of Health (MOH) tracks ``defaults" and ``loss-to-followup" (LTFU) - patients who have missed visits by more than 7 days and 90 days, respectively \citep{Oluoch2021}. At the clinical level, our motivating electronic health record (EHR) data are from The Academic Model Providing Access to Healthcare (AMPATH) care program. AMPATH is a consortium of clinics that treats patients living with HIV in western Kenya. These data contain information on patient-level covariates across several visits over time, information on the scheduled time of each visit as well as the actual time the patient returned for that visit, and other information about death and censoring events. First, our collaborative partners at AMPATH have strong interest in predictive models to help guide outreach efforts for patients deemed to have low probability of retention. Second, there is interest in deploying these models at the point-of-care to recommend ``optimal'' scheduling decisions. The work presented here is motivated by these two goals.

Prediction models can be useful for such problems as they can be used to predict retention rates across different scheduling options. An ``optimal'' scheduling decision can be chosen to be the option that has the highest probability of retention. However, training such models with EHR is complicated for several reasons. First, while we observe the actual return-time for some subjects, other subjects may be transferred out of care before their scheduled visit. This means return-time outcomes are subject to right-censoring. Second, death presents a competing risk problem as patients who die cannot return for their next visit. Third, information on key covariates, such as CD4 count and viral load, is also incomplete due to sporadic monitoring at visits. Finally, optimal scheduling is an inherently causal task that necessitates comparing retention probability under different hypothetical scheduling decisions and recommending the one with the highest potential retention probability. Such comparisons are subject to confounding since, in EHR data, patients are not randomized to, say, return in 2 weeks versus 12 weeks. Rather, patient-level covariates drive both the scheduling decision and the outcome, which make such comparisons prone to confounding. Causal inference procedures are needed to adjust for such observed confounding.

The current state of prediction models in the HIV/AIDS literature has many gaps when addressing these complexities. Many ad-hoc shortcuts are used in practice. For instance, when predicting return-time, many studies artifically right-censor patients at their death time \citep{Ramachandran2020, ridgway2022, Mtisi2023}. This inappropriately indicates that a subject's return-time is greater than their death time. In fact, since death is terminal, we know this patient will never return and so should be removed from the risk set. Regarding right-censoring, certain studies exclude \citep{fentie2022,Nshimirimana2022} ``patients with incomplete information for outcome variable (lost to follow up, dead, drop, transfer out)'' \citep{fentie2022}. Another common alternative is to assume that patients censored before the scheduled time of the next visit will not be retained. This corresponds to an implicit imputation which has been shown to significantly underestimate retention \citep{fox2018, Bengtson2020}. In fact, right-censoring of return time may or may not induce missingness in the retention outcome depending on the definition of retention being used, which we clarify in this paper. A variety of methods for addressing missing covariate information - ranging from the missing-indicator-method (MiM) \citep{Ramachandran2020, Mtisi2023}, to imputation \citep{olatosi2021} - are implemented in practice. The former MiM method consists of including the missingness indicator for each covariate with missing values as a predictor in the model. In this work, we present a causally principled way of accounting for covariate missingness. Finally, since standard ML models are more amenable to binary classification problems, many choose to work with a dichotomous version of the underlying return-time distribution (e.g. missed the visit/did not miss the visit). This shortcut comes at the price of obvious efficiency-loss which our approach avoids. There is also a gap in the methodological literature when it comes to these complexities. Methods for semiparametric causal inference for event-time outcomes have been developed \citep{tsiatis2020}. Bayesian extensions of these models were used to optimize periodontal scheduling \citep{Guan2020}, but not in settings with death, censoring, and missing covariate information. Similarly, some identification results  \citep{rosenbaum1984,Ding2019} and methods \cite{Mitra2016} for causal inference in the presence of missing covariate data have been explored, but remain undeveloped for censored event-time models and so are not applicable to our setting. 

We address these gaps in the following ways. We begin by formalizing and identifying a novel class of causal estimands contrasting retention outcomes under different hypothetical (potentially counterfactual) scheduling decisions. We show that under standard assumptions about the scheduling mechanism, censoring mechanisms, and covariate missingness mechanism, these estimands can be identified in terms of an observable distribution of the waiting time and type of the subsequent event (either return-time or death time). This pair defines the ``state'' of a subject in continuous-time under a hypothetical scheduling decision. Semiparametric Bayesian transition modeling is used to estimate the instantaneous probability of transitioning between states. A post-hoc g-computation procedure simulates this transition process for different hypothetical scheduling decisions to compute posterior retention probabilities. We outline strategies for posterior prediction and optimization for guiding scheduling and outreach interventions. Unlike existing prediction models, patients who die or are censored contribute to the likelihood appropriately. Moreover, as opposed to dichotomizing the underlying return-time outcome, this approach avoids efficiency-loss by modeling it directly in continuous-time. In the following sections we begin by describing the AMPATH EHR data structure for our problem, before defining the relevant estimands and Bayesian transition models. Having established a benchmark approach for retention prediction in the face of several complexities with these transition models, we end the paper with a discussion of the various previously mentioned shortcuts in the HIV/AIDS literature. We explore when they may be defensible and when they can be problematic.

\section{AMPATH Data Structure and Notation} \label{sc:obsdata}

\begin{figure}
     \begin{subfigure}[b]{0.52\textwidth}
          \centering
          \resizebox{.95\linewidth}{!}{\begin{tikzpicture}

\draw (-1.5,.5) node[below, yshift=-.6em]{Subject 1};
\draw (0,0pt) -- (7,0pt);

\draw (0,3pt) -- (0,-3pt) node[below, xshift = 10pt]{$V_1=0$};
\draw (3,3pt) -- (3,-3pt) node[below]{$V_2$};
\draw (4,3pt) -- (4,-3pt) node[below]{$V_3$};
\draw (6,3pt) -- (6,-3pt) node[below]{$V_4$};

\draw[-{Rays[scale=2]} ] (7,3pt) node[below, yshift=-.5em]{$T$};

\draw [decorate,decoration={brace,amplitude=5pt,raise=2ex}]
 (0,0) -- (2.98,0) node[midway,yshift=2em]{$W_1$};
\draw [decorate,decoration={brace,amplitude=5pt,raise=2ex}]
 (3.02,0) -- (3.98,0) node[midway,yshift=2em]{$W_2$};
\draw [decorate,decoration={brace,amplitude=5pt,raise=2ex}]
 (4.02,0) -- (5.98,0) node[midway,yshift=2em]{$W_3$};
\draw [decorate,decoration={brace,amplitude=5pt,raise=2ex}]
 (6.02,0) -- (6.98,0) node[midway,yshift=2em]{$W_4$};

\draw [decorate,decoration={mirror,brace,amplitude=5pt,raise=5ex}]
 (0,0) -- (2.5,0) node[midway,yshift=-3.5em]{$S_1$};
\draw (2.5,3pt) -- (2.5,-3pt) node[below]{};

\draw [decorate,decoration={mirror,brace,amplitude=5pt,raise=5ex}]
 (3,0) -- (3.5,0) node[midway,yshift=-3.5em]{$S_2$};
\draw (3.5,3pt) -- (3.5,-3pt) node[below]{};

\draw [decorate,decoration={mirror,brace,amplitude=5pt,raise=5ex}]
 (4,0) -- (5.5,0) node[midway,yshift=-3.5em]{$S_3$};
\draw (5.5,3pt) -- (5.5,-3pt) node[below]{};

\draw [decorate,decoration={mirror,brace,amplitude=5pt,raise=5ex}]
 (6,0) -- (7.5,0) node[midway,yshift=-3.5em]{$S_4$};
\draw (7.5,3pt) -- (7.5,-3pt) node[below]{};

\draw (-1.5,-2.5) node{Subject 2};
\draw (0,-2.5) -- (5.9,-2.5);
\draw (0,-2.4) -- (0,-2.6) node[below,xshift = 10pt,yshift=-5pt]{$V_1=0$};
\draw (2,-2.4) -- (2,-2.6) node[below, yshift=-5pt]{$V_2$};
\draw (3,-2.4) -- (3,-2.6) node[below, yshift=-5pt]{$V_3$};
\draw[-{Circle[open, scale=2]} ] (6,-2.4) node[below, yshift=-.6em]{$C$};

\draw [decorate,decoration={brace,amplitude=5pt,raise=2ex}]
 (0,-2.5) -- (1.98,-2.5) node[midway,yshift=2em]{$W_1$};
\draw [decorate,decoration={brace,amplitude=5pt,raise=2ex}]
 (2.02,-2.5) -- (2.98,-2.5) node[midway,yshift=2em]{$W_2$};
\draw [decorate,decoration={brace,amplitude=5pt,raise=2ex}]
 (3.02,-2.5) -- (5.98,-2.5) node[midway,yshift=2em]{$W_3$};

\draw [decorate,decoration={mirror,brace,amplitude=5pt,raise=5ex}]
 (0,-2.7) -- (2.3,-2.7) node[midway,yshift=-3.5em]{$S_1$};
\draw (2.3,-2.4) -- (2.3,-2.6) node[below, yshift=-5pt]{};

\draw [decorate,decoration={mirror,brace,amplitude=5pt,raise=5ex}]
(2,-2.9) -- (3,-2.9) node[midway,yshift=-3.5em]{$S_2$};
\draw (2.3,-2.4) -- (2.3,-2.6) node[below, yshift=-5pt]{};

\draw [decorate,decoration={mirror,brace,amplitude=5pt,raise=5ex}]
(3,-2.7) -- (4,-2.7) node[midway,yshift=-3.5em]{$S_3$};
\draw (4,-2.4) -- (4,-2.6) node[below, yshift=-5pt]{};

\end{tikzpicture}}  
          \caption{}
          \label{fig:dagA}
     \end{subfigure}
     \begin{subfigure}[b]{0.47\textwidth}
          \centering
          \resizebox{.95\linewidth}{!}{\begin{tikzpicture}

\draw (-1.5,.5) node[below, yshift=-.6em]{$Y_k(\Delta)= \ ?$};
\draw (0,0pt) -- (3,0pt);

\draw (0,3pt) -- (0,-3pt) node[below, xshift = 0pt]{$V_k$};
\draw[-{Circle[open, scale=2]} ] (3.1,3pt) node[below, yshift=-.6em]{$C$};

\draw [decorate,decoration={brace,amplitude=5pt,raise=2ex}]
 (0,0) -- (2.98,0) node[midway,yshift=2em]{$W_k$};

\draw [decorate,decoration={mirror,brace,amplitude=5pt,raise=5ex}]
 (0,0) -- (2.4,0) node[midway,yshift=-3.5em]{$S_k$};
\draw (2.5,3pt) -- (2.5,-3pt) node[below]{};

\draw [decorate,decoration={mirror,brace,amplitude=5pt,raise=5ex}]
 (2.6,0) -- (4,0) node[midway,yshift=-3.5em]{$\Delta$};
\draw (2.5,3pt) -- (2.5,-3pt) node[below]{};

\draw (-1.5,-2.5) node{$Y_k(\Delta)=0$};
\draw (0,-2.5) -- (4.9,-2.5);
\draw (0,-2.4) -- (0,-2.6) node[below,xshift = 0pt,yshift=-5pt]{$V_k$};
\draw[-{Circle[open, scale=2]} ] (5,-2.4) node[below, yshift=-.6em]{$C$};

\draw [decorate,decoration={brace,amplitude=5pt,raise=2ex}]
 (0,-2.5) -- (4.9,-2.5) node[midway,yshift=2em]{$W_k$};

\draw [decorate,decoration={mirror,brace,amplitude=5pt,raise=5ex}]
 (0,-2.7) -- (2.3,-2.7) node[midway,yshift=-3.5em]{$S_k$};
\draw (2.3,-2.4) -- (2.3,-2.6) node[below, yshift=-5pt]{};

\draw [decorate,decoration={mirror,brace,amplitude=5pt,raise=5ex}]
 (2.4,-2.7) -- (4,-2.7) node[midway,yshift=-3.5em]{$\Delta$};

\end{tikzpicture}}  
          \caption{}
          \label{fig:dagB}
     \end{subfigure}
     \caption{(a)\small  \ Some possible patient trajectories in our data. Time zero (start of followup) is at time of the initial enrollment visit at $V_1=0$. At the enrollment visit, available history is used to schedule a return visit $S_1$ weeks later (measured from $V_1$). The actual waiting time until return is $W_1$. Notice subject 1 returns for visit 2 $W_1$ weeks after $V_1$, at time $V_2$. This is after their scheduled return time, $S_1$, thus they are delayed by $W_1-S_1$ weeks. At time $V_2$, their third visit is scheduled $S_2$ weeks later. Again, they are late and arrive $W_2$ weeks later at time $V_3$. This proceeds until a subject dies at time $T$, before their next scheduled return time. Subject 2, on the other hand returns for their second visit $W_1$ weeks after their initial visit, at time $V_2$. They returned earlier than their scheduled return time of $S_1$ and thus their delay time $W_1-S_1$ is negative. \label{fig:diag}. At their third visit, Subject 2 was scheduled to return in $S_3$ weeks, but was censored at time $C$ (either because they were transferred out of care or due to end of data cut). (b) $\Delta$-Retention, defined as $Y_k(\Delta) = I( W_k - S_k < \Delta, \delta=1)$, is missing for subjects censored before time $V_k+S_k+\Delta$, but observed for subjects censored after that time.}
 \end{figure}
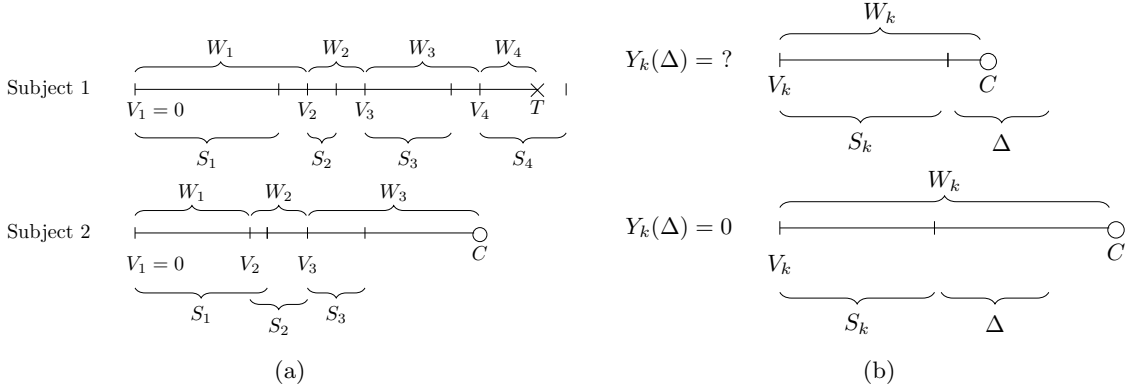
We use a subset of the AMPATH EHR data consisting of subjects diagnosed with HIV who are newly enrolled at an AMPATH clinic between 2014 and 2023. For each subject, we observe data on $j=1, 2,\dots, J$ visits with $j=1$ being their first visit after enrolling in AMPATH and $J$ itself is subject-specific. The $j^{th}$ visit occurs at time $V_j>0$. Throughout, we will measure time in weeks. Time zero is considered to be the time of their first visit, $V_1$. At visit $j$, information may be collected on a $P-$dimensional covariate vector $\tilde L_j=(\tilde L_{1,j}, \tilde L_{2,j},\dots, \tilde L_{P,j})$. In reality, not every covariate will be measured at each visit. For instance, in the AMPATH EHR, viral load and CD4 count are sometimes monitored and, therefore, available while at other visits  they are not monitored and therefore unavailable. Let $M_j = (M_{1j}, M_{2j}, \dots, M_{Pj})$ be a collection of indicators at visit $j$, where $M_{p,j}=1$ indicates that $\tilde L_{pj}$ was monitored and $M_{p,j}=0$ indicates that it was not monitored. We denote the observed set of covariates at visit $j$ as $ L_j = \tilde L_j \odot M_j$, where $\odot$ denotes element-wise multiplication. A zero in $L_j$ indicates a missing value if the corresponding element of $M_j$ is 0. Formally, we should write $L_j^{M_j}$, which denotes the vector of observed covariate values in monitoring pattern $M_j$, but we avoid this indexing by $M_j$ throughout for compactness. Similarly, we let $ L_j^U = \tilde L_j \odot (1-M_j)$ denote the unobserved values at visit $j$. At visit $j$, the covariate information available at this visit is encoded in the pair $( L_j, M_j)$. The information in the previous $j-1$ visits consists of the previously monitored covariates $(\bar{L}_{j-1}, \bar M_{j-1})$, where the notation  $\bar X_j$ denotes history, $\bar X_j=(X_{1}, X_2, \dots, X_j)$. We will also denote $S_j$ as the scheduled return time at visit $j$, measured in weeks from current visit at time $V_j$. For instance, $S_j=2$ indicates that, at visit $j$ occurring at time $V_j$, the patient was scheduled to return in two weeks.  Thus, in addition to covariates, at visit $j$ we also have access to the scheduled return times of the previous $j-1$ visits, $\bar S_{j-1}$. In clinical practice, typically the scheduling options are integers such as $S_j\in \mathcal{S} = \{2,4,8\}$ weeks. After visit $j$ is over, there is some waiting time until the next event $W_j = \min ( W_{Vj}, W_{Tj}, W_{Cj})$. Here, $W_{Vj} = V_{j+1} - V_j$, $W_{Tj} = T-V_j$, $W_{Cj} = C - V_j$ are the waiting times until next visit, death, and censoring from time $V_j$, respectively. The corresponding indicator $\delta_j \in\{1, 0, -1\}$ indicates a waiting time until subsequent visit, death, or censoring event - whichever comes first. A few example trajectories are depicted in Figure \ref{fig:dagA}. The available information at visit $j$ just ahead of the scheduling decision for the next visit, $S_j$, includes the covariates monitored at that visit along with information from previous visits $ H_j = \{ \bar{L}_{j}, \bar M_{j}, \bar S_{j-1}, \bar W_{j-1}, \bar \delta_{j-1}=1\}$. In the AMPATH EHR, censoring was either due to end of the data cut at July 31, 2023 or when a patient was transferred out of AMPATH care - whichever came first. The observed data consists of $\mathcal{D} = \{ \bar{L}_{J_i}, \bar M_{J_i}, \bar S_{J_i}, \bar W_{J_i}, \bar \delta_{J_i}\}_{i=1}^n$, where $J_i$ indicates the total number of events observed for subject $i$. 

\subsection{$\Delta-$Retention, Censoring, and Competing Events}

Formally, we define retention as a dichotomization of the underlying return-time outcome. Specifically, let $Y_j(\Delta) = I( W_j - S_j \leq \Delta, \delta_j = 1)$ indicate a return visit within $\Delta>0$ weeks of a scheduled visit or earlier. We call this ``$\Delta$-retention'' as it formally defines the concept of retention in the presence of competing and censoring events. Since $Y_j(\Delta)$ is just a transformation of $(W_j, \delta_j)$, we exclude it from $\mathcal{D}$ to avoid redundancy. The specified constant $\Delta$ is a tolerance threshold for a delay in the scheduled visit which can be set depending on the outcome of interest. For example, the Kenyan Ministry of Health (MOH) HIV treatment guidelines describes a patient as lost to followup (LTFU) if they do not return within 90 days of their last scheduled visit". Thus, $\Delta = 90/7 \approx 13$ weeks if we want to avoid LTFU. Kenyan MOH guidelines also define a ``defaulter" as a patient who does not return past seven days, which corresponds to $\Delta=1$. Finally, interruption in treatment (IIT) is another key outcome and is defined as missing a visit by more than 4 weeks, corresponding to $\Delta=4$ for avoiding IIT. 

The desired outcome after visit $j$ is $Y_j(\Delta)=1$. If $Y_j(\Delta)=0$, then either the patient returned late by more than $\Delta$, was transferred from care sometime after $\Delta$, or died - all of which are undesirable. Note that if a subject was censored \textit{before} $\Delta$, then $Y_j(\Delta)$ is missing because we do not know whether they returned for a visit within $\Delta$ at another clinic. On the other hand, if this subject were censored some time $\Delta$ weeks after their scheduled return time, then $Y_j(\Delta)=0$. This is because, even though we never observe their return time, we know they at least did not return within $\Delta$ weeks of their scheduled visit. This is illustrated in Figure \ref{fig:dagB}. Two common approaches for dealing with this missingness is to 1) exclude them from the analysis \citep{fentie2022, Nshimirimana2022} or 2) labeling them as not having been retained, which results in underestimating retention \citep{Bengtson2020}. The first approach inefficiently excludes subjects with missing data. The latter corresponds to a type of unintentional, implicit imputation. Unlike censoring which may render $Y_j(\Delta)$ missing, death give us information about $Y_j(\Delta)$. Specifically, if a patient's first event after visit $j$ is a death event, then we know for sure that $Y_j(\Delta)=0$ since death precludes a subsequent return visit. In contrast to these shortcuts, our approach allows both censored and dead patients to contribute information about $Y_j(\Delta)$ appropriately.

\section{Causal $\Delta$-Retention and Identification} \label{sc:pos}

Optimizing scheduling decisions is necessarily a causal endeavour as it involves predicting retention outcomes under counterfactual scheduling decisions and choosing the scheduling option that maximizes the probability of retention. Due to lack of randomization, estimation with EHR data is subject to confounding: healthy subjects with, say, low measured viral load may be scheduled further out and, at the same time, may be more likely to return on time. To disentangle effects of the scheduling decision from such confounding factors we take a causal approach and begin by defining potential event-time outcomes. For those at risk of a subsequent event at visit $j$, let $W_j^s = \min(W_{Vj}^s, W_{Tj}^s)$ denote the potential waiting time from $V_{j}$ until the next event, had we scheduled a patient to come back $s\in\mathcal{S}$ weeks later. It is the first of either the potential waiting time until the next visit, $W_{Vj}^s$, or the potential waiting time until death, $W_{Tj}^s$. Note we should write $s_j$, but omit the subscript for compactness since $Y$ and $W$ already include subscript $j$. Let the potential event type be $\delta_j^s = I(W_{Vj}^s < W_{Tj}^s )$. Let $ Y^s_{j}(\Delta) = I( W^s_{j} - s \leq \Delta, \delta_{j}^s = 1) $ denote the potential $\Delta$-retention. Among the subgroup of patients at-risk of an event at visit $j$ with available history $H_j = ( \bar L_j, \bar M_j, \bar S_{j-1}, \bar W_{j-1}, \bar \delta_{j-1}=1)$, we are interested in estimating the potential retention probability
\begin{equation} \label{eq:poprob}
    \Psi_j^s(H_j;\Delta) = P(Y^s_j(\Delta) = 1 \mid \bar L_j, \bar M_j, \bar S_{j-1}, \bar W_{j-1}, \bar \delta_{j-1}=1) \ \ \ \text{ for } s \in\mathcal{S}
\end{equation}
This is the proportion of the target sub-population, defined by strata of $(\bar L_j, \bar M_j, \bar S_{j-1}, \bar W_{j-1})$, at risk for a subsequent return visit (i.e. $\bar \delta_{j-1}=1$) that would have been $\Delta-$retained in care had we made scheduling decision $s$. Formally, $\Psi_j^s(H_j;\Delta)$ is a ``heterogeneous'' treatment effect, because it allows for effect of $s$ on retention to vary across subgroups defined by $H_j$.  If we had this function (or an estimate of it), then for a given patient at visit $j$ we could plug in their history and obtain their retention probability. This can be used post-scheduling to flag patients with low probability of retention under their scheduled return time. Additionally, pre-scheduling, we would like to recommend an optimal scheduling decision, $s_j^*(h_j)$, in the sense that $s_j^*(h_j) = \underset{s\in\mathcal{S}}{\argmax } \ \Psi_j^s(h_j;\Delta) $. That is, for someone with history $h_j$, we make a scheduling decision $s_j^*(h_j)$ by comparing counterfactual retention probabilities within the subpopulation defined by $h_j$ under different scheduling decisions and choosing the one that maximizes retention probability. Since $Y_j^s(\Delta)$ is a function of $W_j^s$ and $\delta_j^s$, identification of the quantities $ \Psi_j^s(H_j;\Delta)$ and $s_j^*(h_j)$ requires identifying the distribution of potential outcomes $f^*_{jk}(W^s_j=w, \delta^s_j = k \mid H_j=h_j)$ in terms of the observed data distribution. A key identification assumption is that the scheduling mechanism must be conditionally ignorable: Among those at-risk at visit of a subsequent event after visit $j$ (i.e. $\bar \delta_{j-1}=1$), the \textit{monitored} covariate information, $(\bar{L}_{j}, \bar M_{j})$, and previous scheduling and waiting time history, $(\bar S_{j-1}, \bar W_{j-1})$, are sufficient to control for confounding between the scheduling decision at the current visit and the waiting time until the next event. Formally, $W_{Vj}^s, W_{Tj}^s \perp S_j \mid  \bar S_{j-1}, \bar W_{j-1},  \bar{L}_{j}, \bar M_{j} =\bar m_j, \bar \delta_{j-1}=1$. Such an assumption has been invoked before with partially missing covariates \citep{rosenbaum1984, Ding2019,Kreif2020} and is defensible if non-monitored covariate values are unknown to the physician and, therefore, could not have influenced the scheduling decision. This is the case in our setting, where viral load and CD4 counts are missing because a test was not run and, therefore, the results are unknown to the clinician at the time of scheduling. Since the other assumptions are variations of the usual causal assumptions such as positivity, SUTVA, and non-informative censoring, we enumerate them in the supplement. Under these assumptions, we can identify $f_{jk}^*(w, \delta_j^s=k \mid H_j=h_j)$ in terms of the observed data sub-density functions of visit and death time \citep{tsiatis2020}, 
$$ f_{jk}^*(W^s_j=w, \delta^s_j=k \mid H_j=h_j) = dw^{-1 }\lim_{dw \rightarrow 0 } P( w \leq W_j < w + dw, \delta_j = k \mid S_j = s, H_j =h_j)$$ 
for event type $k\in \{0,1\}$. Note the right-hand side is the observed data sub-density function of the waiting time, $W_j$, from visit $j$ to an event of type $k$: $f_{jk}(w \mid S_j=s, H_j=h_j)=dw^{-1 }\lim_{dw \rightarrow 0 } P( w \leq W_j < w + dw, \delta = k \mid S_j = s, H_j = h_j)$. From a transition modeling perspective, it is the instantaneous probability of transitioning from state $\delta_{j-1} = 1$ (current visit) to state $\delta_j=k$ at time $w$. From a competing risk perspective, it reflects the fact that death competes with return to be the first event after visit $j$. The various quantities of interest $ \Psi_j^s(H_j;\Delta)$  are merely integrals over this distribution. For instance, 

$$  \Psi_j^s(h_j;\Delta) = \sum_{k\in\{0,1\}}\int_0^\infty I( w - s < \Delta, k=1 ) f_{jk}(w \mid S_j=s, H_j=h_j)d w  $$

In the subsequent sections, we will outline Bayesian semiparametric estimation of the cause-specific hazards and a g-computation procedure for conducting posterior inference on functionals such as $\Psi_j^s(h_j;\Delta)$ and $s_j^*(h_j)$.

\section{Flexible Bayesian Transition Models for Retention} \label{sc:mods}

It is known that the sub-density of each event can be expressed completely in terms of the cause-specific hazards of each event type \citep{kalbfleisch2011}. That is, for some conditioning set, abbreviated $\bullet$ for compactness, $f_{jk}(w\mid \bullet) =  \lambda_{jk}(w \mid \bullet) \exp\big( - \int_{0}^{w} \sum_{k\in\{0,1\} }\lambda_{jk}(u \mid  \bullet ) du \big) $
Where $\lambda_{jk}(w \mid  \bullet) = dw^{-1} \lim_{dw\rightarrow 0}  P(w \leq W_j < w + dw, \delta_j = k \mid W_j>w,\bullet)$ is the cause-specific hazard. A posterior over the cause-specific hazards of the two events, $\lambda_{jk}(w \mid  \bullet )$ for $k\in \{0,1\}$, induces a posterior on $f_{jk}(w\mid \bullet)$. To that end, we propose a stratified proportional hazard model for the hazard of an event of type $k$ after visit $j$. Let $X_{j}$ denote covariates and let $Z_j$ be a discrete variable across which we can stratify the model. As is commonly done in causal models in time-varying settings, additionally invoke a Markov assumption and condition on information in the history $H_j$ only up through the previous $(j-1)^{th}$ visit. We will stratify by each scheduling time-monitoring pattern combination so that $Z_j$ denotes each unique level of $(S_j=s, M_j= m_j)$ and $X_{j} =  (L_j,  W_{j-1}, S_{j-1})$ indicates the collection of the observed confounders in pattern $ M_j= m_j$, $L_j$, actual return-time of the previous visit, $W_{j-1}$, and scheduled return-time of the previous visit $S_{j-1}$, respectively. Inclusion of the last two is important since an excellent predictor of retention following visit $j$ is often the delay time of the previous visit $W_{j-1} - S_{j-1}$. The stratified proportional hazard model can then be written as
\begin{equation}
    \lambda_{jk}(w \mid X_j=x_j, Z_j=z) = \mu_{jkz}(w) \exp( x_j' \beta_{jkz} )
\end{equation}
Each stratum has its own baseline hazard $\mu_{jkz}$ and multiplicative effects of covariates governed by $\beta_{jkz}$. That is, we use the patients at-risk  of a subsequent event at visit $j$ to fit separate models for each scheduling-monitoring pattern combination, with each model conditioning on covariates observed in that monitoring pattern. Thus the dimensionality of $\beta_{jkz}$ and $x_j$ vary across strata $z$, will be no larger than $P+2$. We will have a flexible specification of the former and keep a linear specification of the covariate effects, though tools such as b-splines can be used to incorporate non-linear functions of continuous components of $x_j$. Throughout, we place $N(0,3)$ priors on the coefficients $\beta_{jkz}$, which is wide on the exponentiated (hazard ratio) scale. As a concrete example in our data, at the first visit $j=1$, there are two possible covariates of interest that have missingness - CD4 count and viral load. This yields four possible monitoring patterns - those with both missing, both observed, and those missing one or the other. There are two possible subsequent events - death $(k=0)$ and the next visit ($k=1$). And we are interested in three possible scheduling options, $\mathcal{S} = \{2,4,8\}$. This leads to two hazard models (one for each event type) for each of the $4\cdot3=12$ scheduling-monitoring strata.
\begin{figure}[h!]
    \centering
    \includegraphics[scale=.45]{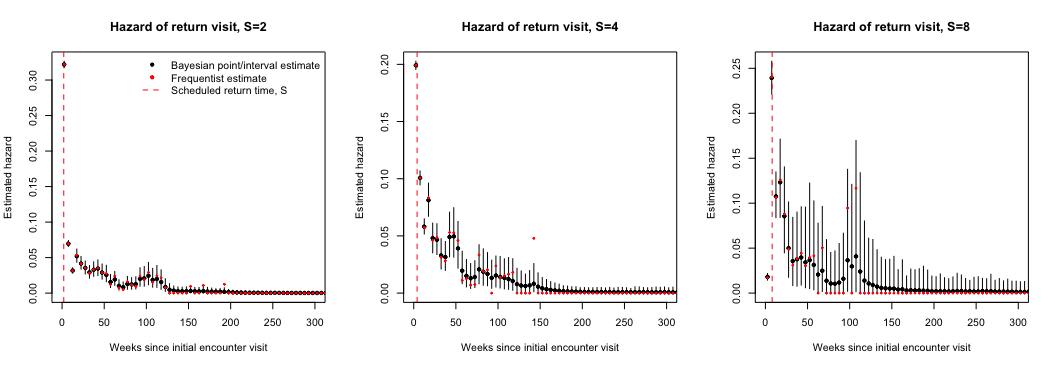}
    \caption{Unadjusted estimates of the hazard of return time from visit $j=1$, stratified by scheduled time. The Bayesian model is able to capture clumping around the scheduled visit time as seen by the spike in the hazard estimate at the dashed vertical line. Finally, as seen from weeks 50-150 in the second panel and around week 100 in the third panel, the gAR1 prior smooths the hazard estimates even as the frequentist point estimates are jumpy due to low numbers of patients at risk.}
    \label{fig:haz_plots_example}
\end{figure}
For the baseline hazard in each stratum, we employ a flexible piece-wise constant specification of the form $ \mu_{jkz}(w) =  \sum_{u=1}^U I(w \in \mathcal{T}_{u,jkz} ) \theta_{u,jkz} $. 
The collection $\{\mathcal{T}_{u,jkz}\}_{u=1}^U$ is a set of $U$ non-overlapping, equal-width intervals that partition the followup time window from 0 to the maximum observed waiting time in each stratum. The parameters $\{ \theta_{u,jkz}\}_{u=1}^U$ are a collection baseline hazard rates, where  $\theta_{u,jkz}$ is the hazard rate in interval $\mathcal{T}_{u,jkz}$. A finer partition (larger $U$) leads to a more flexible specification that can capture baseline hazards with clumping at the scheduled time. To enforce smoothness and, therefore, reduce sensitivity to choice of partition, we specify a first-order autoregressive Gaussian prior process on the log hazard rates. \textit{A priori}, the interval-specific log hazard rate is given by $ \log \theta_{1,jkz} = \eta + \sigma \epsilon_1$. For $u=2,3,\dots, U$, we set
 $\log \theta_{u,jkz} = \eta(1-\rho) + \rho \log \theta_{u-1,jkz} + \sigma \epsilon_u $. Here, $0\leq\rho<1$, $-\infty <\eta <\infty$, and $\sigma>0$ are the hyperparameters of the process and $\epsilon_u$ are i.i.d. $N(0,1)$ random variables. We denote this as $\{ \theta_{u,jkz} \}_{u=1}^U \sim gAR1(\rho, \eta, \sigma)$. Note that the prior mean of this process on the log scale is $E[\log  \theta_{u,jkz}] = \eta$, shrinking towards a constant-hazard baseline hazard in the absence of data. Additionally, we allow correlation between the hazards, since under this prior $Corr(\log \theta_{u,jkz}, \log\theta_{u-v,jkz}) = \rho^v$. This reflects the prior belief that, in the absence of data at interval $u$, the hazard rate in that interval is similar to hazard rates in recent intervals. This adds stability to the baseline hazard rate estimation at later time points with low at-risk counts. We specify wide priors for the hyperparameters $\rho$, $\eta$, and $\sigma$. Figure \ref{fig:haz_plots_example} depicts unadjusted cause-specific hazards for subsequent return after their initial visit $j=1$. It can be seen that this estimator is both flexible enough to capture complexities such as clumping at the scheduled return-time while being smoother than the empirical counterpart, in red.

\section{Posterior Retention Prediction and Optimization} \label{sc:sampling}
The posterior distribution for the unknown hazard rates $\{ \theta_{u, jkz} \}_{u=1}^U$ and coefficients $\beta_{jkz}$ are not available in closed form under the autoregressive prior outlined in the last section. Instead, we use Hamiltonian Monte Carlo (HMC) to obtain posterior draws of these parameters. We developed a user-friendly \texttt{R} package, \texttt{causalBETA} (see \cite{ji2023} for a vignette), that can be used to return draws from the posterior of the parameters in the described hazard models. Thus, after a suitable burn-in period, we have available $a=1,2,\dots, A$ posterior draws of the parameters, denoted $\{ \theta_{u, jkz}^{(a)} \}_{u=1}^U$ and coefficients $\beta_{jkz}^{(a)}$ for each $j$, $k$, and $z$. The former set of draws is used to construct a draw of the baseline hazards via $\mu_{jkz}^{(a)}(w) = \sum_{u=1}^U I(w \in \mathcal{T}_{u,jkz} ) \theta_{u,jkz}^{(a)} $. Consider a subject currently at visit $j$ with monitoring pattern $m_{j} $ and scheduled return time in $s_{j}$ weeks. The latter two factors determines their stratum membership, $Z_j=z_{j}$. Their observed covariates are $x_{j} =  (L_{j},  W_{j-1}, S_{j-1})$. Prediction and optimization for this subject can be done as follows. First, the $a^{th}$ draw of this subject's retention probability can be obtained via $b=1,2,\dots, B$ Monte Carlo simulations as follows:
\begin{enumerate}
    \item Simulate waiting time until subsequent visit: $W_{Vj}^{(b)} \sim \lambda_{j1}(w \mid x_{j}, z_{j}) = \mu_{j1z_{j}}^{(a)}(w) \exp\Big( x_{j}' \beta_{j1z_{j}}^{(a)} \Big)$. 
    This can be done using the discrete inverse-CDF method.
    \item Simulate waiting time until death: $ W_{Tj}^{(b)} \sim \lambda_{j0}(w \mid x_{j}, z_{j}) = \mu_{j0z_{j}}^{(a)}(w) \exp\Big( x_{j}' \beta_{j0z_{j}}^{(a)} \Big) $
    \item Compute the time and type of the first event, given by $W_j^{(b)} = \min( W_{Vj}^{(b)},  W_{Tj}^{(b)})$ and $\delta^{(b)}_j = I( W_{Vj}^{(b)} <  W_{Tj}^{(b)})$, respectively. Then, approximate
    $$ \Psi_j^s(H_j;\Delta)^{(a)} \approx \frac{1}{B} \sum_{b=1}^B I\Big(W_j^{(b)} - s_{j} < \Delta, \delta_{j}^{(b)} = 1 \Big) $$
\end{enumerate}
Repeating this for each posterior draw $a=1,2,\dots, A$ we have a set of posterior draws $ \{ \Psi_j^s(H_j;\Delta)^{(a)}\}_{a=1}^A$ which can be used to make inferences about the subject's retention probability $\Psi_j^s(H_j;\Delta)$. A point estimate can be found by taking the median while a 95\% credible interval can be found by taking the 2.5th and 97.5th percentiles of the draws. 

For a given subject with covariate $x_j$, optimization across many scheduling options $s\in\mathcal{S}$ can also be done. In our application we consider $\mathcal{S} = \{2,4,8\}$, which are typical in clinical practice. For each posterior draw, repeat steps 1-3 above under each $s \in \mathcal{S}$ and obtain $\{ \Psi_j^s(H_j;\Delta)^{(a)}\}_{s\in\mathcal{S}}$. The $a^{th}$ posterior draw of the optimal scheduling option, $s_j^*(h_j)^{(a)}$, is the option with the highest $\Psi_j^s(H_j;\Delta)^{(a)}$ acrosss $s$. Doing this for all $A$ gives a set of posterior draws of the optimal option $ \{ s_j^*(h_j)^{(a)} \}_{a=1}^A$, which can be used to make inferences about $s_j^*(h_j)$. Across posterior draws, we can estimate an entire posterior probability mass function for the optimal decisions, 
\begin{equation}
    P( s_j^*(h_j) = k \mid \mathcal{D}) \approx \frac{1}{M} \sum_{m=1}^M I(s_j^*(h_j)^{(a)}=k) 
\end{equation}
for each $k\in \mathcal{S}$. The option with the highest posterior probability of being the optimal rule, $\hat s_j^*(h_j)$,  is used as a point estimate of the unknown optimal decision. However, the entire distribution can be used to quantify our uncertainty around this point estimate, as we will demonstrate in our data application. For three scheduling options, $k=\{2,4,8\}$, the vector $\{P( s_j^*(h_j) = 2 \mid \mathcal{D}), P( s_j^*(h_j) = 4 \mid \mathcal{D}), P( s_j^*(h_j) = 8 \mid \mathcal{D})\}$ lives in a $3-$simplex and uncertainty around the posterior mode,  $\hat s_j^*(h_j)$, can be visualized on a machina triangle, as will be illustrated in the application.

\section{Simulation Experiments} \label{sc:sims}
The main methodological contribution of this work is development of principled estimands and prediction strategies with missing outcomes and covariates regardless of the choice of model. However, a comparison of models can be useful as the quality of the predictions and optimization with the proposed transition models has not been evaluated in this setting. In this section, we assess performance relative to Bayesian Additive Regression Trees (BART) \citep{chipman2010} trained on a dichotomous retention outcome, $Y_j(\Delta)$, where values missing due to censoring are implicitly imputed as not retained. This is to mimic machine learning (ML) risk prediction models popular in HIV/AIDS. The dichotomization is done because common ML packages easily accommodate binary classification tasks, but do not accommodate continuous outcomes with complexities of censoring and competing events. Of course, this comes at the expense of information loss. It also mimics naive implicit imputations done in the presence of censoring, which has led to underestimation of retention \citep{Bengtson2020}. Such approaches lead to poor predictions regardless of the flexibility of the downstream predictive modeling.

Without loss of generality, we set $j=1$ so that we are analyzing retention following the first visit. To assess performance, we simulate a training set with $500$ subjects each in the following way. For each subjects we simulate six covariates (three binary and three continuous). Of those six covariates, two covariates (one binary and one continuous) can potentially be missing in ways that are depending on the other 4 covariates. Conditional on observed covariates, one of three possible scheduling times are assigned to each subject. Conditional on covariates and a scheduled return time, a death time, and censoring time are simulated from separate Weibull hazard models. To mimic clumping around the scheduled time, a subsequent return-time is simulated from a two-part mixture of a point-mass at the scheduled time and a covariate-dependent Weibull distribution. The next event is the first of the three simulated events. In the same way, we simulate a test set with $500$ subjects where there is no censoring (i.e. censoring time is set to infinity) so that we have complete outcome information $Y_1(\Delta) \in\{0,1\}$. In this simulation, we set $\Delta=2$ to analyze two-week delay. In each iteration, we use the training set to estimate the described models, and compute posterior mean of retention probability under the assigned scheduled return time, $\Psi_1^s(H_1;\Delta)$, for each subject in the test set. We use average test-set AUC to summarize predictive performance across simulated datasets, which are presented in Panel A of Table \ref{tab:sim_res_auc}.

\begin{table}[h!]
\centering
\begin{tabular}{lllllll}
 \multicolumn{7}{c}{Panel A: Average Test-Set AUC} \\ \hline
 & \multicolumn{3}{c}{Low Censoring} & \multicolumn{3}{c}{High Censoring} \\ \hline
\multicolumn{1}{c|}{Covariate Missingness}  & Logistic  & BART      & \multicolumn{1}{l|}{BTM} & Logistic & BART & BTM \\ \hline
\multicolumn{1}{c|}{None}                   & .644      & .642      &  \multicolumn{1}{l|}{.697} & .561  & .557  &  .669 \\
\multicolumn{1}{c|}{Low}                    & .583      & .570      &  \multicolumn{1}{l|}{.635} & .551  & .526  &  .617 \\
\multicolumn{1}{c|}{High}                   & .580      & .570      &  \multicolumn{1}{l|}{.631} & .544  & .527  &  .610 \\ \hline
\multicolumn{7}{c}{} \\
\multicolumn{7}{c}{Panel B: Accuracy of Estimated Optimal Scheduling Time } \\ \hline
 & \multicolumn{3}{c}{Low Censoring} & \multicolumn{3}{c}{High Censoring} \\ \hline
\multicolumn{1}{c|}{Covariate Missingness}  & Logistic  & BART & \multicolumn{1}{l|}{BTM} & Logistic & BART & BTM \\ \hline
\multicolumn{1}{c|}{None}                   & .578      & .574 & \multicolumn{1}{l|}{.786} & .305 & .316  & .659  \\
\multicolumn{1}{c|}{Low}                    & .479      & .475 & \multicolumn{1}{l|}{.598} & .369 & .351  & .524 \\
\multicolumn{1}{c|}{High}                   & .468      & .489 & \multicolumn{1}{l|}{.593} & .358 & .352  & .516 \\ \hline
\end{tabular}
\caption{Panel A: Average test-set area under the curve (AUCs) of predicted retention probabilities for various missingness/censoring scenarios. Panel B: Average proportion of test-set patients for whom the method correctly identified the optimal scheduling decision. Results averaged across 300 simulation runs each with $n=500$ subjects in both panels.}
\label{tab:sim_res_auc}
\end{table} 
Results labeled ``BTN'' is the developed Bayesian transition model while ``BART'' is the described tree-based model. We also include maximum likelihood estimates of $\Psi_1^s(H_1;\Delta)$ from a logistic regression of dichotomous $Y_1(\Delta)$ on all covariates using the training set - labeled ``Logistic''. In order to satisfy conditional ignorability in the presence of missingness, for all methods, we stratify the models based on missingness pattern. That is, we fit a separate model in each of the four possible missingness patterns, each adjusting for covariates observed in that pattern. A prediction is generated for a test set subject using the model that corresponds to that subject's missingness pattern. Panel A of Table 1 presents AUCs under a low and high censoring in the training set. These have censoring rates approximately 20\% and  40\% on average across simulation runs. In addition, we simulate across a range of covariate missingness rates. The setting ``None'' is the simplest setting where all covariates are observed. In the ``Low'' setting, about 45\% of subjects have at least one of the two partially observed covariates missing. In the ``High'' setting, about 60\% have at least one of the two covariates missing. Note that since the return-time is generated from a two-component mixture, all models are misspecified. Thus, across scenarios, what is more important is the relative test-set AUC, rather than the absolute AUC. We see that the proposed approach has higher AUC than BART. The performance differential is maintained even as censoring rates increase. This is likely because our method works with the underlying continuous return-time and allows censored and dead patients to contribute to the model through their respective cause-specific hazards. In contrast, the other methods rely on dichotomous return-time.

In addition to test-set predictive performance under an observed scheduling time, we are also interested in how well the method can estimate the optimal scheduling time $s_1^*(h_1)$. For each subject in the test set, we compute the true scheduling decision (out of the possible 3) that maximizes $\Psi_1^s(H_1;\Delta)$ using that subject's covariates and true model parameters. We then check to see if this matches the mode of $s_1^*(h_1)$ across posterior draws. In Table 1 Panel B, we see BTM is able to accurately estimate the optimal scheduling decision for 78.6\% of test set subjects, on average, across simulation runs in the low-censoring-no-covariate-missingness setting. Performance dips to 59.8\% and 59.3\% with low and high covariate missingness, respectively. The accuracy of the optimal estimate obtained by comparing estimates of $\Psi_1^s(H_1;\Delta)$ under various $s$ using BART and logistic regression all perform worse. In the setting with high censoring, BART and logistic regression exhibit worse performance across covariate missingness levels. Again, the loss of information and inability to have censored and dead patients contribute fully to the model is aggravated when censoring rates are high. One conclusion here is that a full generative, semiparametric model for the return time yields adequate estimates of the causal quantities of interest across these censoring and missingness settings and outperforms common off-the-shelf alternatives. In cases where there is very minimal or no censoring, then $Y_1(\Delta)$ is observed for everyone. In this case, the difference between stratified BTM and stratified BART will be negligible.

\section{Modeling HIV Care Retention in AMPATH} \label{sc:analysis}

We used the developed method to analyze retention following the first visit $(j=1)$ using electronic health records from AMPATH. This is because patients are believed to be at highest risk of disengagement at the beginning of care. Patient data were collected from four clinics - Kitale, Busia, Uasin Gishu District Hospital (UGDH), and Kitale County Referral Hospital Module A (KCRH). We include all adults with HIV with their first visit occurring between between January 1, 2014 and and July 31, 2023 (end of the data cut), which totals to $n=11,011$. All patients have recorded a scheduled return time, gender, age, whether or not they have ever been on antiretrovirals (ARV), and the particular clinic they visited. Additionally, we have viral load and CD4 count recorded for patients who have had these values monitored. For patients who have not, this information is missing. A summary of the data is given in Table \ref{tab:sumstats}. Across clinics, a plurality of patients have both CD4 count and viral load missing, but a significant subgroup have viral load observed while having CD4 count missing. A very small fraction have both measured. The most common scheduling decision was a two-week return time across sites. There were some notable baseline differences in ARV and age across clinics.

\begin{table}[h!]
\centering\small
\begin{tabular}{lcccc}
 & \multicolumn{4}{c}{Clinic} \\ \cline{2-5} 
 & \multicolumn{1}{c}{Kitale} & \multicolumn{1}{c}{Busia} & \multicolumn{1}{c}{ UGDH} & \multicolumn{1}{c}{KCRH Module A} \\
 & (n=3398) & (n=2891) & (n=2438) & (n=2284) \\ \hline
Missing: &  &  &  &  \\
\hspace{.1in} viral load only & 1389 (.41) & 607 (.21) & 303 (.12) & 895 (.39) \\
\hspace{.1in} CD4 only & 93 (.03) & 114 (.04) & 50 (.02) & 73 (.03) \\
\hspace{.1in} both & 1907 (.56) & 2167 (.75) & 2077 (.85) & 1313 (.57) \\
\hspace{.1in} neither & 9 (\textless{}.01) & 3 (\textless{}.01) & 8 (\textless{}.01) & 3 (\textless{}.01) \\
Observed log viral load & 7.7 (4.8-10.1) & 8.0 (5.7-10.6) & 7.5 ( 4.7-9.9) & 7.1 (5.0-7.7) \\
Observed log CD4 & 5.2 (4.6-6.1) & 5.6 (5.1-6.3) & 5.4 (4.9-6.2) & 5.2 (4.8-6.0) \\
Age at enrollment & 36.1 ( 28.0-42.4) & 36.0 ( 28.0-42.7) & 35.0 (26.8-41.7) & 40.2 (31.8-47.5) \\
Male & 1261 (0.37) & 1064 (0.37) & 895 (0.37) & 890 (0.39) \\
On antiretrovirals (ARV) & 2710 (0.80) & 2688 (0.93) & 2169 (0.89) & 2265 (0.99) \\ 
Scheduled return time &  &  &  &  \\
\hspace{.1in} Two weeks & 2848 (0.84) & 2099 (0.73) & 2004 (0.82) & 1899 (0.83) \\
\hspace{.1in} Four weeks. & 485 (0.14) & 671 (0.23) & 354 (0.15) & 348 (0.15) \\
\hspace{.1in} Eight weeks & 65 (0.02) & 121 (0.04) & 80 (0.03) & 37 (0.02) \\ \hline
\end{tabular}
\caption{Summary Statistics for patients at their first visit after enrollment. Mean and Interquartile range (IQR) presented for continuous features. Count and proportions presented for discrete features. Statistics are stratified by clinic at which visit was made. Note there is substantial variability in features across clinic in terms of scheduling decisions (e.g. Busia) and ARV (e.g. Kitale has especially lower rates) and age at enrollment (e.g. KCRH having older patients). Abbreviations: UGDH -Uasin Gishu District Hospital; KCRH- KCRH Module A.  }
\label{tab:sumstats}
\end{table}

\begin{figure}[h!]
    \centering
    \includegraphics[scale=.35]{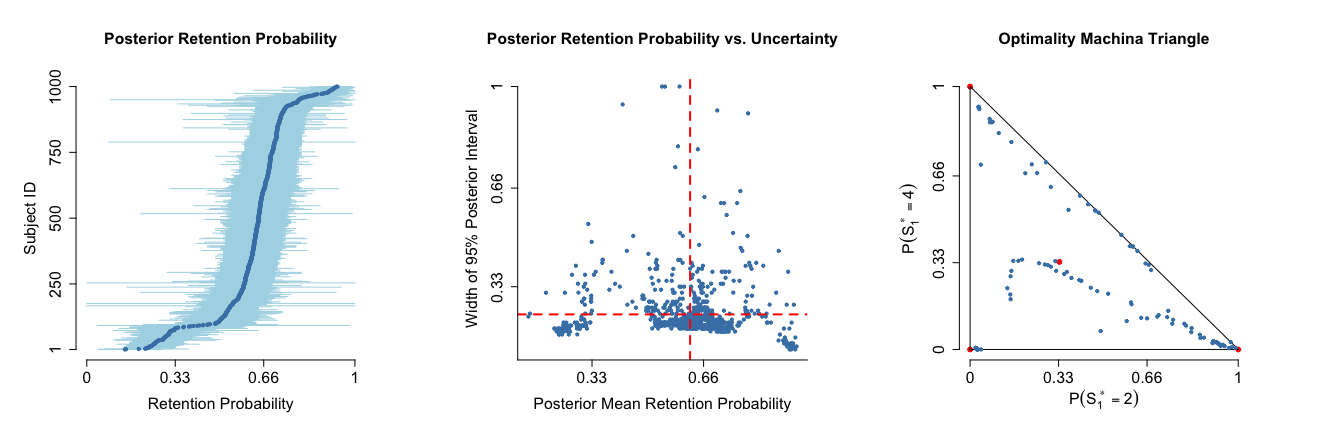}
    \caption{For all plots, $j=1$ and $\Delta=90/7\approx13$. Left: Posterior mean and 95\% credible interval of $\Psi_j^s(h_{ji};\Delta)$ for each subject $i$ in the test set under their scheduled return time $s=s_i$. Middle: For each subject, this plot displays the posterior mean of $\Psi_j^s(h_{ji};\Delta)$ on the x-axis along with with the width of the credible interval of $\Psi_j^s(h_{ji}$. Right: Each subject's posterior PMF of the optimal scheduling time representing in a machina triangle. $ P( s_j^*(h_j) = 2 \mid \mathcal{D})$ is on the x-axis and $ P( s_j^*(h_j) = 4 \mid \mathcal{D})$ on the y-axis. The red point in the middle represents maximal uncertainty with each option having 1/3 posterior probability of being the optimal option. Patients closest to vertex $(0,0)$, $(0,1)$, and $(1,0)$ are those with option 8, 4, and 2 as the posterior mode optimal rule, respectively.}
    \label{fig:analysis_plot}
\end{figure}

We focus our analysis on loss to follow-up (LTFU) by defining retention with $\Delta=90/7$. We randomly select $1,000$ subjects from the full data set to be in a test set that is held out from model training. The remaining $10,011$ subjects are used for training. We fit cause-specific hazards described in Section \ref{sc:mods} stratified by each clinic, scheduling decision, and missingness pattern combination. In each missingness pattern, we adjust for covariates observed in that pattern. In patterns where either viral load or CD4 are measured, they are included in the modeling using b-splines to allow for non-linear functional forms in the cause-specific hazard models. A key advantage of the Bayesian approach is that the posterior draws of the model parameters need only be obtained once and stored. Quantities of interest can then by computed on-demand by post-processing these draws - making it ideal for running at point-of-care dashboards. This is in contrast to frequentist procedures that may require complex bootstrapping procedures be run for each new quantity of interest. The left panel of Figure \ref{fig:analysis_plot}, for instance, displays posterior mean and 95\% credible interval for each subject's retention probability. Some subject are predicted to be highly likely to be retained, but with high uncertainty. This is partially because some strata are very small. For instance, the return time models among patients in Busia who were scheduled to return in eight weeks was trained on fewer than 121 subjects (see Table \ref{tab:sumstats}), with the precise number depending on the monitoring pattern. Thus, predictions for test set subjects in this stratum have larger posterior uncertainty. A helpful way of making decisions under uncertainty is shown in the middle panel of Figure \ref{fig:analysis_plot} which plots the width of the 95\% credible interval against the posterior mean for each subject's retention probability. The red dashed lines at the empirical averages of these values divides the space into four quadrants, which can be used to guide outreach decisions. For instance, the patients in the bottom-left quadrant are ones who we are relatively more certain have a lower than average posterior probability of being retained. They are likely excellent candidates for outreach efforts. In contrast, the patients in the top-left quadrant also have a lower-than-average posterior probability of being retained. But given the higher posterior uncertainty, if outreach resources are scare they may be lower priority.

The right panel of Figure \ref{fig:analysis_plot} conveniently displays the posterior distribution of the optimal scheduling decision for each subject as described in Section \ref{sc:sampling}. The posterior mass function over the optimal scheduling decision for a subject is given by the triple $\{P( s_j^*(h_j) = 2 \mid \mathcal{D}), P( s_j^*(h_j) = 4 \mid \mathcal{D}), P( s_j^*(h_j) = 8 \mid \mathcal{D})\}$, which can be represented as a point in this triangle. The red point represents maximal uncertainty about which of the option maximizes $\Delta$-retention, giving uniform 1/3 posterior probability to each option. The red points on each of the three vertices represents high posterior certainty that the corresponding option is the optimal one. For instance, points near the vertex $(0,0)$ are patients with $\approx 1$ posterior probability of $S=8$ being the optimal decision. These diagrams can be used to make decisions under uncertainty in the following way. Consider two patients who are above the line $y=1/3$ on the triangle, but one is closer to the vertex $(0,1)$ than the other. For both of these patients, the posterior mode optimal rule is $\hat s_j^*(h_j)=4$. However, the one closer to $(0,1)$ is the one for whom we have higher posterior precision around this mode. Thus, if we had only one appointment time slot 4 weeks from the current visit, we may consider scheduling the patient closer to $(0,1)$.

\begin{figure}[h!]
    \centering
    \includegraphics[scale=.30]{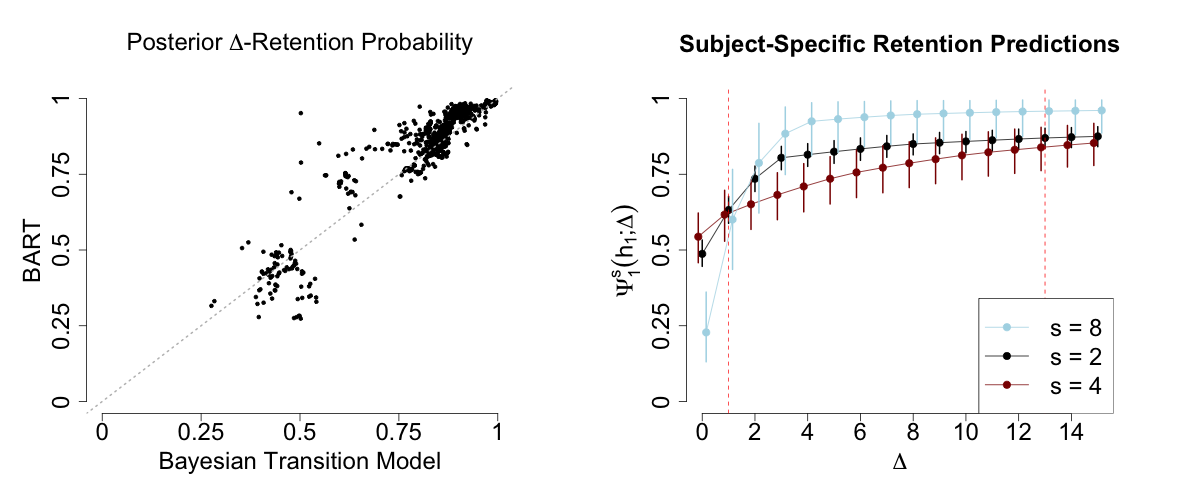}
    \caption{Left: For $j=1$ and $\Delta=90/7$, each point represents $\Psi_j^s(h_{ji};\Delta)$ for each subject $i$ in the test set under their scheduled return time $s=s_i$ under BART vs. the formulated Bayesian Transition Model. Due to low censoring rates, the predictions produced by the two approaches are quite consistent, as shown by the scattering of points around the 45-degree line. Right: For one subject, $i$, in the test set, BMT produced estimates of  $\Psi_1^s(h_{1i};\Delta)$ for various $\Delta$ under each possible scheduling decision $s\in\{2,4,8\}$. This plot illustrates that ability of our Bayesian Transition Model to produce point and interval estimates under a variety of retention definitions, $\Delta$, and scheduling options without need to re-run the models. It illustrates that optimization for a single $\Delta$, while dominant in the HIV/AIDS literature, is not global and motivates future work on optimization via more general utility functions.}
    \label{fig:prediction_plot}
\end{figure}

We also ran prediction models using BART for LTFU. Since censoring and death rates were quite low in the data ($\approx 1\%$ and $\approx 10\%$, respectively) the predictions were quite similar when we correctly flag dead patients as not being retained and simply drop censored subjects. This can be seen in the left panel of Figure \ref{fig:prediction_plot}. Each point is a subject in the test set and the scattering around the 45 degree line depicts general agreement in the predictions. But, as shown in the simulation, higher censoring rates may lead to differences in performance. Lastly, rather than keeping focus at a particular $\Delta=90/7\approx 13$ weeks, we could consider estimation retention, $\Psi^s_1(h_1;\Delta)$ across $\Delta$. The right panel of Figure \ref{fig:prediction_plot} depicts this for a single subject. Essentially this traces out the posterior subdistribution function \citep{kalbfleisch2011} of the delay time, $W_1 - s_1$, under three different scheduling decisions ($s\in\{2,4,8\}$). The significance of this plot is to show that optimization for a given $\Delta$ does not yield a decision that is globally optimal across all $\Delta$. For this subject, scheduling the next visit in 8 weeks maximizes retention under LTFU definition of $\Delta\approx 13$ (indicated by the dashed vertical line). However, this minimizes retention under the default definition ($\Delta=1$). Global optimality would only occur if the counterfactual subdistribution function of delay time under decision $s=8$ was above the the subdistribution functions under the other decisions across all $\Delta$. 
Thus, while vast majority of clinical and applied work in HIV retention modeling focuses on a single $\Delta$, our analyses suggests optimizing based on the entire subdistribution function could be desirable area of future work. This can perhaps be done by seeing which decision has the largest area under the subdistribution function or optimizing based on an expected utility function that takes a weighted average of specified points on the subdistribution function.

\section{Discussion and Connections to Common Shortcuts}

In terms of clinical significance of the work, our initial results with AMPATH data presented here suggest that we can identify a subgroup of patients that have both high probability and certainty of missing a subsequent visit and, therefore, are strong candidates for outreach interventions, such as a reminder phone call. This enables data-driven use of scarce outreach resources at AMPATH. An eventual product of this work is to use the models developed here to display recommended scheduling times on a clinician's dashboard at the point-of-care, given their inputted patient features. 

In terms of methodological significance, we proposed a novel $\Delta$-retention estimand that provides a formal causal target of inference for retention prediction problems in HIV/AIDs for the first time. We proposed a Bayesian method for estimating these quantities in the presence of a competing death event, right-censoring, and missingness in the covariate space. It is worth explicitly connecting our approach with the often-implemented shortcuts discussed in the introduction. First, here we explicitly clarify that right-censoring of return-time $W_{Vj}$ may or may not induce missingness in $Y_j(\Delta)$, depending on when censoring occurs relative to the chosen $\Delta$. Dropping subjects with missing outcomes \citep{fentie2022, Nshimirimana2022} will yield a training set that may not be representative of a target population or test set. Assuming that these subjects did not return \citep{ridgway2022} is a type of ``worst-case'' imputation where we set $Y_j(\Delta)=0$ for those with a missing outcome. This may or may not be defensible depending on the problem. We also clarify that rather than treating $Y_j(\Delta)$ as censored at death, the competing nature of death implies that $Y_j(\Delta)=0$. As in our data application, we show that such shortcuts may not make a material difference on estimation when censoring rates are low. In our simulation section, we show that differences can be material when censoring rates are high. 

A second point of clarification is regarding missingness in covariates, which is routine in EHR. Our analysis is explicitly centered around decision-making/intervening on the scheduling process. When covariates are missing because they were not monitored, imputation is not necessary if ignorability holds conditional on monitored covariates. This is perhaps the underlying assumption behind MiM approaches common in this literature \citep{Ramachandran2020, Mtisi2023}. Including a missingness indicator as a covariate in the model may be seen as a smoother alternative to our full stratification across the missingness pattern. This smoother alternative may be justified if some missingness patterns are sparse.

\newpage

\section*{Acknowledgements}
This work was partially funded by NIH Award R01AI167694 and PCORI Award ME2023C131348.

\bibliography{refs} 

\newpage

{\centerline{\Large\textbf{Supplement}} }

\vspace{.25in}

{\large\textbf{S1. Identification Assumptions}}

The main identification result in the paper follows from the following set of assumptions on the censoring and and scheduling mechanisms.

\begin{enumerate}
    \item \textbf{Conditionally ignorable scheduling mechanism}: Among those at risk at visit of a subsequent event after visit $j$ (i.e. $\bar \delta_{j-1}=1$), the monitored covariate information, $(\bar{L}_{j}, \bar M_{j})$, and previous scheduling and waiting time history, $(\bar S_{j-1}, \bar W_{j-1})$, are sufficient to control for confounding between the scheduling decision at the current visit and the waiting time until the next event.
    $$ W_{Vj}^s, W_{Tj}^s \perp S_j \mid   \bar{L}_{j}, \bar S_{j-1}, \bar W_{j-1}, \bar M_{j} =\bar m_j, \bar \delta_{j-1}=1 $$
     Note that this implies that conditioning on the monitored covariates is sufficient to control for confounding and has been invoked before \citep{rosenbaum1984}. This is valid in settings, such as ours, where non-monitored covariate values are unknown to the physician and, therefore, could not have influenced the scheduling decision. Assuming satisfy this condition
    \item \textbf{Scheduling positivity}: For all $h_j$ such that $f_{H_j}(h_j) >0$, each scheduling decision $s\in \mathcal{S}$ must be possible: $P(S_j =s \mid H_j =h_j ) > 0$.
    \item \textbf{Non-informative censoring}: Among those at risk for a subsequent event at visit $j$, the cause-specific hazard of being censored before the next event is at that time is unrelated to potential outcomes conditional on available history at visit $j$
    \begin{equation*}
        \begin{split}
                   & \lim_{dw \rightarrow 0} \frac{ P(w \leq W_j < w + dw, \delta_j  = -1  \mid W_j>w, H_j=h_j, W^s_{Vj}, W^s_{Tj} )}{dw}  \\
                                    & \ \ \ \ \ = \lim_{dw \rightarrow 0}  \frac{ P(w \leq W_j < w + dw, \delta_j = -1 \mid W_j>w, H_j=h_j)}{dw}
        \end{split}
    \end{equation*}
    This ensures that, conditional on history, censored subjects are not systematically more likely to, say, die before their next visit.
    \item \textbf{Censoring positivity}: For each $h_j$ such that $f_{H_j}(h_j) >0 $ and possible event time $w$, it must be possible to remain uncensored long enough to have the next event, $P( W_{Cj} > w \mid H_j = h_j ) > 0$. If this did not hold, then there would be some subgroup who would always be censored and we would have no information on their potential outcomes.
    \item \textbf{SUTVA}: $W_{Cj} = \sum_{s\in\mathcal{S}} I(S=s)W_{Cj}^s$, $W_{Vj} = \sum_{s\in\mathcal{S}} I(S=s)W_{Vj}^s $, and $W_{Tj} = \sum_{s\in\mathcal{S}} I(S=s)W_{Tj}^s $
\end{enumerate}

\newpage
{\large\textbf{S2. Likelihood and Posterior Inference}}

Here we present the likelihood contribution of subject $i$ with $J_i$ events. $J_{i-1}$ of those events are visits and the $J_i^{th}$ event is either death ($k=1$) or censoring event ($k=-1$). Thus, the likelihood is,
\begin{equation*}
    \begin{split}
        f( \bar L_{iJ_i}, \bar M_{iJ_i}, \bar S_{iJ_i}, \bar W_{iJ_i}, \bar \delta_{iJ_{i-1}}=1, \delta_{iJ_i}=k ) & \propto f_{J_ik}(w_{iJ_i}, \delta_{iJ_i}=k \mid \bar S_{iJ_i}, \bar L_{iJ_i}, \bar M_{iJ_i}, \bar W_{iJ_{i-1}},\bar \delta_{iJ_{i-1}}=1 ) \\
        & \ \ \times \prod_{j=1}^{J_i-1} f_{j1}(w_{ij}, \delta_{ij}=1 \mid \bar S_{ij}, \bar L_{ij}, \bar M_{ij}, \bar W_{ij-1}, \bar \delta_{j-1}=1 )  \\
    \end{split}
\end{equation*}
The subdensify functions, $f_{jk}$, above can be expressed in terms of the cause-specific hazards as:
$$ f_{jk}(w\mid- ) =  \lambda_{jk}(w \mid  - ) \exp\Big( - \int_{0}^{w} \sum_{k\in\{0,1\} }\lambda_{jk}(u \mid  - ) du \Big) $$
Where for compactness $-$ is a placeholder for the conditioning set. Thus, subjects contribute information about hazard of a subsequent return visit for each of their observed return visits. If they die after their last recorded return visit, the contribute information about the hazard of a subsequent death event after that visit. If they are censored after their last recorded visit ($k=-1$), then the information that death and return visit are greater than that censoring time via $ \exp\Big( - \int_{0}^{w} \sum_{k\in\{0,1\} }\lambda_{J_ik}(u \mid  S_{J_i}=s, H_{J_i}=h_{J_i} )$. It should be understood that $\lambda_{jk}$ are functions of the unknown parameters $\omega_{jk} = \Big\{ \beta_{jkz}, \{ \theta_{u,jkz}\}_{u=1}^U\Big\}$. We will denote the full set of parameters as $\omega = \{\omega_{j1}, \omega_{j0} \}_{j=1}^{\max_i J_i}$ . The full likelihood then is, 

$$ \mathcal{L}(\mathcal{D} \mid \omega ) = \prod_{i=1}^n  f( \bar L_{iJ_i}, \bar M_{iJ_i}, \bar S_{iJ_i}, \bar W_{iJ_i}, \bar \delta_{iJ_i} )$$

Depending on the visit of interest, not all contributions for a subject are required. For instance, inference for $\Psi_1^s(h_1;\Delta)$ only requires drawing from the posterior of the parameters governing $f_{1k}(w\mid- )$. Thus, only contributions from subjects at risk for event ($j=1$) is needed. Similarly, if inference is required for $j=2$, we only require drawing from the posterior of the parameters governing $f_{2k}(w\mid- )$. This likelihood can be specified in standard Bayesian software such as \texttt{Stan} along with the priors used in the paper to obtain draws from the corresponding posterior.

\end{document}